\documentclass[aps,twocolumn,groupedaddress,superscriptaddress,amsmath,amssymb,prb,longbibliography]{revtex4-2}
\usepackage{mathtools,amsmath,amsxtra}
\usepackage{amssymb}
\usepackage{physics}
\usepackage[english]{babel}
\usepackage{dsfont}
\usepackage{graphicx}
\usepackage{dcolumn}
\usepackage{bm}
\usepackage{bbold}
\usepackage[usenames,dvipsnames]{color}
\usepackage[colorlinks,citecolor=Blue,linkcolor=Red, urlcolor=Blue]{hyperref}
\usepackage{tikz}
\usetikzlibrary{decorations.pathmorphing}
\usetikzlibrary{arrows.meta}
\usepackage{braket}
\usepackage[normalem]{ulem}
\usepackage{lipsum}
\usepackage{caption}
\DeclareCaptionType{equ}[][]

\begin{document}

\title{Many-Body Models for Chirality-Induced Spin Selectivity in Electron Transfer}

\author{A. Chiesa}
\affiliation{Universit\`a di Parma, Dipartimento di Scienze Matematiche, Fisiche e Informatiche, I-43124 Parma, Italy} 
\affiliation{INFN–Sezione di Milano-Bicocca, gruppo collegato di Parma, 43124 Parma, Italy}
\affiliation{Consorzio Interuniversitario Nazionale per la Scienza e Tecnologia dei Materiali (INSTM), I-50121 Firenze, Italy}

\author{E. Garlatti}
\affiliation{Universit\`a di Parma, Dipartimento di Scienze Matematiche, Fisiche e Informatiche, I-43124 Parma, Italy} 
\affiliation{INFN–Sezione di Milano-Bicocca, gruppo collegato di Parma, 43124 Parma, Italy}
\affiliation{Consorzio Interuniversitario Nazionale per la Scienza e Tecnologia dei Materiali (INSTM), I-50121 Firenze, Italy}

\author{M. Mezzadri}
\affiliation{Universit\`a di Parma, Dipartimento di Scienze Matematiche, Fisiche e Informatiche, I-43124 Parma, Italy} 
\affiliation{INFN–Sezione di Milano-Bicocca, gruppo collegato di Parma, 43124 Parma, Italy}

\author{L. Celada}
\affiliation{Universit\`a di Parma, Dipartimento di Scienze Matematiche, Fisiche e Informatiche, I-43124 Parma, Italy} 
\affiliation{INFN–Sezione di Milano-Bicocca, gruppo collegato di Parma, 43124 Parma, Italy}

\author{R. Sessoli}
\affiliation{Dipartimento di Chimica “U. Schiff” (DICUS), Universit\`a degli Studi di Firenze, I-50019 Sesto Fiorentino (FI), Italy}
\affiliation{Consorzio Interuniversitario Nazionale per la Scienza e Tecnologia dei Materiali (INSTM), I-50121 Firenze, Italy}

\author{M. R. Wasielewski}
\affiliation{Department of Chemistry, Center for Molecular Quantum Transduction, and Institute for Sustainability and Energy at Northwestern, Northwestern University, Evanston, Illinois 60208-3113, United States}

\author{R. Bittl}
\affiliation{Freie Universität Berlin, Fachbereich Physik, Berlin Joint EPR Lab, D-14195 Berlin, Germany}

\author{P. Santini}
\affiliation{Universit\`a di Parma, Dipartimento di Scienze Matematiche, Fisiche e Informatiche, I-43124 Parma, Italy} 
\affiliation{INFN–Sezione di Milano-Bicocca, gruppo collegato di Parma, 43124 Parma, Italy}
\affiliation{Consorzio Interuniversitario Nazionale per la Scienza e Tecnologia dei Materiali (INSTM), I-50121 Firenze, Italy}

\author{S. Carretta}
\email{stefano.carretta@unipr.it}
\affiliation{Universit\`a di Parma, Dipartimento di Scienze Matematiche, Fisiche e Informatiche, I-43124 Parma, Italy} 
\affiliation{INFN–Sezione di Milano-Bicocca, gruppo collegato di Parma, 43124 Parma, Italy}
\affiliation{Consorzio Interuniversitario Nazionale per la Scienza e Tecnologia dei Materiali (INSTM), I-50121 Firenze, Italy}

\begin{abstract}
We present the first microscopic model for chirality-induced spin selectivity effect in electron-transfer, in which the internal degrees of freedom of the chiral bridge are explicitly included. By exactly solving this model on short chiral chains we demonstrate that a sizable polarization on the acceptor arises from the interplay of coherent and incoherent dynamics, with strong electron-electron correlations yielding many-body states on the bridge as crucial ingredients. Moreover, we include the coherent and incoherent interactions with vibrational modes and show that they can play an important role in determining the long-time polarized state probed in experiments.
\end{abstract}

\maketitle

\twocolumngrid 

The Chirality-Induced Spin Selectivity (CISS) effect is a stunning but still not understood phenomenon in which electrons are strongly spin polarized when passing across chiral molecules.  
It has attracted an increasing interest \cite{Bloom2024,Naaman2019} for explaining chemical reactions \cite{Naaman2020} or biological processes \cite{Bloom2024,Michaeli2016}. Moreover, by inducing a very large electron spin polarization at high temperature and without external fields, CISS was immediately proposed as a tool for room-temperature spintronics \cite{Yang2021} and, more recently, for quantum technologies \cite{Aiello2022,Chiesa2023}. \\
CISS was detected on a variety of systems \cite{Bloom2024} 
and using different experimental techniques \cite{Ray1999,Naaman2020b,Bloom2024}. 
Many models were put forward to explain the widely explored transport and photoemission setups \cite{Alwan2024}, but a univocal, satisfactory explanation is still lacking \cite{evers2022}. 
Spin polarization is triggered by spin-orbit coupling (SOC) \cite{Gutierrez2012,Guo2012,Cuniberti2020}, but this is small in organic systems and hence additional ingredients are needed to amplify the polarization to the level observed experimentally. Single-electron models \cite{Gutierrez2012,Guo2012,Guo2014,Medina2015,Matityahu2016,Pan2016,Varela2016,Michaeli2019} generally yield small polarization in presence of strong decoherence. Different models for transport indicate that this can be enhanced \cite{Alwan2024,Fransson2019,Fransson2020,Fransson2021,Fransson2022,Das2022}  by electron-electron \cite{Fransson2019,Fransson2021} or electron-vibration \cite{Fransson2020,Du2020,Zhang2020,Michaeli2023} interactions.

To disentangle the role of various ingredients, a different approach exploiting a simpler setup was recently proposed \cite{Chiesa2021,Fay2021,Luo2021}. By focusing on photo-induced electron transfer (ET) through a chiral molecular bridge, it is indeed possible to remove complex interfaces or metals with large SOC and focus only on the chiral bridge. It was theoretically shown that an exchange interaction between the moving electron and the unpaired one localized on the donor can lead to a spin polarization \cite{Fay2021b}. However, such interaction is lacking in transport or photoemission experiments. Moreover, this model only focuses on the initial and charge-separated states, without a microscopic description of the internal degrees of freedom of the bridge.

Very recently, CISS was actually observed in photo-induced ET \cite{Eckvahl2023}. This finding definitely moves the focus to the only player present in all  experimental setups, i.e. the chiral bridge, whose explicit description in ET is still lacking \cite{Fay2021,Fay2021b}. 

Here we provide such a microscopic description by a fermionic many-body model explicitly accounting for the internal degrees of freedom of the chiral bridge. The large spin polarization (above 50 \%) observed both in ET \cite{Eckvahl2023} and in transport measurements \cite{Giaconi2023} on short chiral molecules indicates that the underlying mechanism must arise even from a minimal model on a limited number of sites, which therefore is the focus of our work. This allows for an exact numerical solution of the ET dynamics, including electron-electron interactions and both coherent and incoherent coupling with vibrational modes, in contrast with transport models that usually require strong approximations \cite{Fransson2019,Michaeli2023}.
Moreover, such a minimal model provides a deep insight into the polarization mechanism and allows us to explore different regimes of parameters to investigate their role.  \\
We obtain a sizable  polarization with realistic parameters and we highlight the crucial role of electron-electron interactions, consistently 
with models proposed for transport \cite{Fransson2019,Fransson2020,Alwan2024}. 
Finally, we  investigate the long-time relaxation of the donor-chiral bridge-acceptor supramolecule, establishing opposite local polarization on donor and acceptor with the bridge in the ground singlet, as experimentally observed \cite{Eckvahl2023}. \\ 
The combination of these tunable  models with targeted experiments could be the key to fully understand CISS and to harness it for technological applications.
\begin{figure*}[t!]
    \centering
    \includegraphics[width=0.8\textwidth]{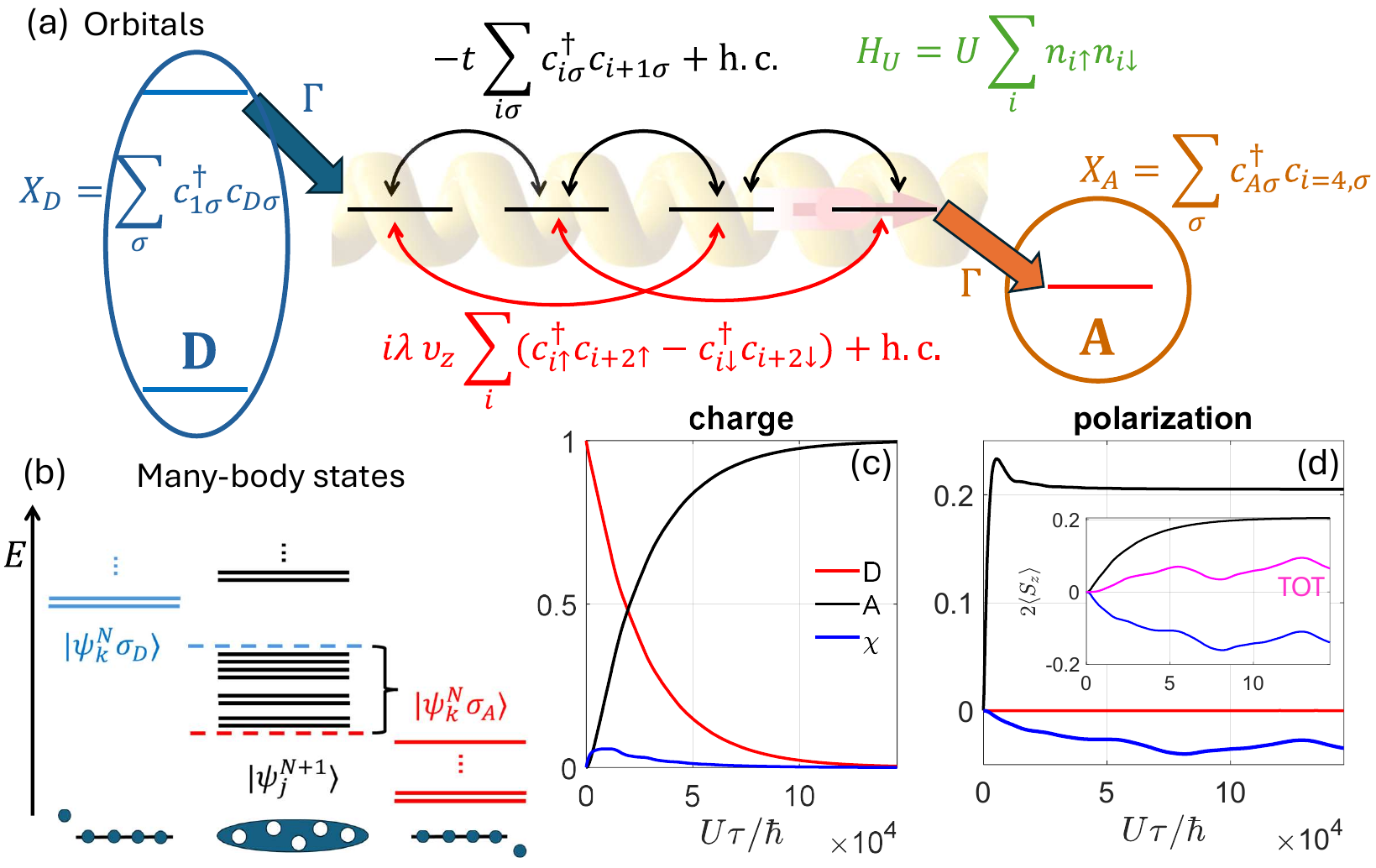}
    \caption{{\bf System description and polarization dynamics}. (a) Model system, consisting of a 4-site chiral bridge with nearest-neighbors spin-independent hopping (black), next-to-nearest-neighbors SOC (red, only the most relevant $z$ component reported) and on-site Coulomb repulsion (green). Sites 1 and 4 are  coupled to the donor (excited) and acceptor states by an incoherent spin-independent hopping driven by a large energy gap, characterized by a rate $\Gamma$. (b) Energy level structure of the bridge many-body states with $N+1$ (black) electrons. In blue (red) the energy levels for $N$ electron on the bridge and one electron on the donor (acceptor).
    Only the lowest block is in the right energy window to be involved in the ET. 
    (c,d) Example of simulated electron transfer dynamics. (c) Charge on D, A and $\chi$ (having subtracted its initial value, 4). (d) Corresponding spin polarization on the acceptor $p_A = (n_{A,\uparrow}-n_{A,\downarrow})/(n_{A,\uparrow}+n_{A,\downarrow})$ 
    (black), the donor (red), 
    and on the bridge $\sum_i(n_{i,\uparrow}-n_{i,\downarrow})$ (blue). Inset: spin expectation values along the bridge axis for the bridge $2\langle S_{z,\chi}\rangle= \langle \sum_{i=1}^4 2 S_{z,i} \rangle$ (blue), the acceptor $2\langle S_{z,A}\rangle$ (black), and the sum of the two (magenta).
    Detailed oscillations of $n_i$ and $p_i$  ($i = 1, \ldots, 4$) are shown in Fig. S2. Simulation parameters: $t/U = 0.0125$, $\lambda/U = 6.25 \times 10^{-4}$, $\Gamma/U= 2.5 \times 10^{-4}$.}
    \label{fig1}
\end{figure*}

{\it Model system --} To investigate the role of correlations, we consider a prototypical many-body model and describe the chiral bridge $\chi$ as a half-filled chain of $N$ orbitals and $N$ electrons with a singlet ground state, characterized by the following Hamiltonian:
\begin{eqnarray} \nonumber
    H_\chi &=& -t \sum_{i=1}^{N-1} \sum_{\sigma} c_{i,\sigma}^\dagger c_{i+1,\sigma} + U \sum_{i=1}^N n_{i \uparrow} n_{i \downarrow} \\
    &+& i \lambda \sum_{i = 1}^{N-2}  \sum_{\sigma\sigma^\prime} c^\dagger_{i,\sigma} \boldsymbol{\upsilon}_i  \cdot \boldsymbol{\sigma} \, c_{i+2,\sigma^\prime} + {\rm h.c.} 
    \label{eq:Ham}
\end{eqnarray}
Here $c_{i\sigma}^\dagger$ ($c_{i\sigma}$) creates (annihilates) an electron with spin $\sigma$ on site $i$ and $n_{i\sigma} = c_{i\sigma}^ \dagger c_{i\sigma}$. The first term is the spin-independent hopping between neighboring sites characterized by $t$, the second is the on-site Coulomb repulsion of strength $U$, and the last one is the spin-orbit coupling in the form of a spin-dependent next-to-nearest neighbor  
hopping with the coupling constant $\lambda$ \footnote{$c^\dagger_{i,\sigma} \boldsymbol{\upsilon}_i  \cdot \boldsymbol{\sigma} \, c_{i+2,\sigma^\prime}$  is a short-hand notation (see, e.g., Ref. \cite{Cuniberti2020}) for $\upsilon_{ix} \left( c^\dagger_{i,\uparrow} c_{i+2,\downarrow} + c^\dagger_{i,\downarrow} c_{i+2,\uparrow} \right) - i \upsilon_{iy} \left( c^\dagger_{i,\uparrow} c_{i+2,\downarrow} - c^\dagger_{i,\downarrow} c_{i+2,\uparrow} \right) + \upsilon_{iz} \left( c^\dagger_{i,\uparrow} c_{i+2,\uparrow} - c^\dagger_{i,\downarrow} c_{i+2,\downarrow} \right)$.}. We follow the minimal model of \cite{Fransson2019}, which does not focus on a specific system and hence yields general qualitative conclusions \footnote{Hamiltonian \eqref{eq:Ham}
ensures the presence of two channels for electron transfer and hence opens the possibility of a spin polarization. Indeed, we have checked that no polarization arises in presence of only nearest-neighbor 
interactions both in the hopping and SOC terms of Eq. \eqref{eq:Ham}, consistent with reports for transport in a two-terminal setup \cite{Guo2012,Guo2014b,Geyer2019}.}.
The vector $\boldsymbol{\upsilon}_i$ is defined referring to a helix shape of the molecule with a single turn, radius $a$, pitch $c$ and positions of the sites ${\bf r}_i = [a \cos (i-1)2\pi/(N-1), a \sin (i-1)2\pi/(N-1), (i-1)c/(N-1)]$, $\boldsymbol{\upsilon}_i = {\bf d}_{i+1} \times {\bf d}_{i+2}$ and ${\bf d}_{i+s} = ({\bf r}_i-{\bf r}_{i+s})/|{\bf r}_i-{\bf r}_{i+s}|$.
Here we assume radius equal to the pitch as in \cite{Michaeli2019}. Changing the enantiomer corresponds to the transformation $\left( \upsilon_{xi}, \upsilon_{yi}, \upsilon_{zi} \right) \rightarrow \left( -\upsilon_{xi}, \upsilon_{yi}, -\upsilon_{zi} \right)$. 
Donor (D) and Acceptor (A) are modeled by two additional sites at the two ends of the bridge [see Fig. \ref{fig1}-(a) for $N=4$], incoherently coupled to it (see below). A fast incoherent multi-step ET was directly observed 
in many systems, such as the PXX-NMI$_2$-NDI molecule where CISS was recently shown \cite{Eckvahl2023}. 
In Eq. \eqref{eq:Ham} we have assumed degenerate orbitals, but this condition can be relaxed, as discussed below.
In addition, we have not explicitly taken into account the electrostatic contribution to the energy due to the charge separation from D to A \cite{Ratner2013} \footnote{This effect is implicitly included in the hierarchy of the many-body states.}. \\
\begin{figure*}[t!]
    \centering
    \includegraphics[width=0.85\textwidth]{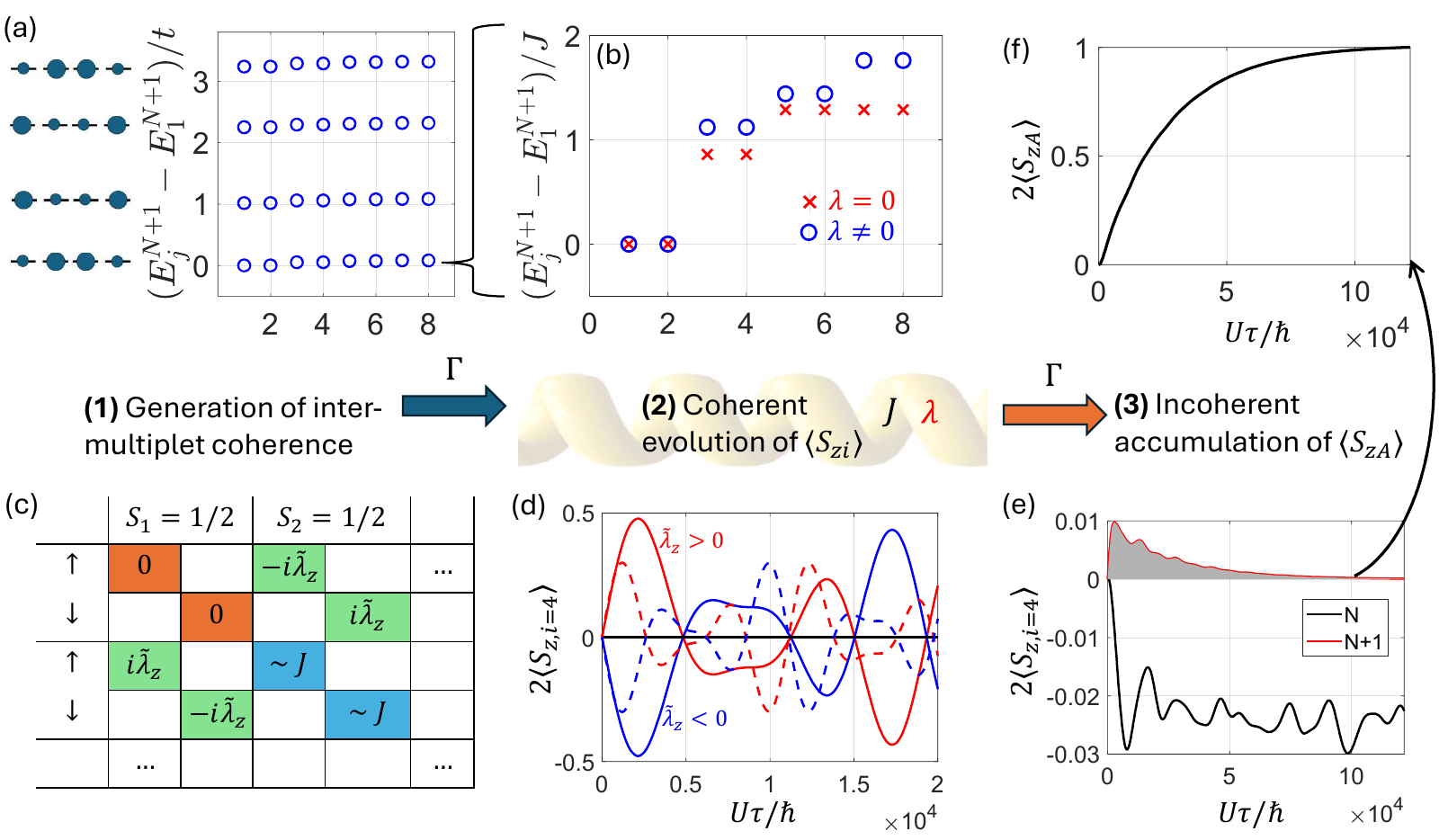}
    \caption{{\bf Polarization mechanism.}
    (a) Many-body states $\ket{\psi_j^{N+1}}$ ($j=1,\ldots,32$) within the $(N+1)$-electron subspace with a single double occupation (with charge distribution sketched on the left) and (b) zoom on the lowest energy block. 
    (c) Scheme of populations and coherences on the density matrix generated by the jump operator $X_D$ from D to $\chi$ and corresponding matrix elements of the Hamiltonian (explicitly shown for two lowest energy doublets of panel (b) with $\tilde{\lambda}_z \propto \lambda \upsilon_z$). (d) Hamiltonian time evolution of $2\langle S_{z,i=4}\rangle$ for an initial state prepared into $X_D \rho(0) X_D^\dagger$ (with population only in the lowest energy block of Fig. \ref{fig2}-(a), i.e. $\ket{\psi_j^{N+1}}$, $j = 1,...,8$) and using $t/U = 0.0125$, $\lambda_z/U = 0.0005$. Results for the two enantiomers (corresponding to an opposite value of $\tilde{\lambda}_z$) are represented by different colors, while dashed lines are obtained by halving correlations (i.e. using $t/U = 0.025$, $\lambda_z/U = 0.001$). (e) Full time evolution of $2\langle S_{z,i=4}\rangle$ including both coherent and incoherent dynamics in Eq. \eqref{eq:Redfield} and separated into the two contributions of the states with either $N$ or $N+1$ electrons on $\chi$. The latter 
    is proportional to the derivative of $S_{zA}$ accumulated on the acceptor (f), which can therefore be obtained from the shaded area in (e).}
    \label{fig2}
\end{figure*}
We describe the dynamics after photo-excitation, initially with two electrons on the two relevant orbitals on the donor [blue oval in Fig. \ref{fig1}-(a)] in a singlet state. In the absence of a coherent coupling between D and $\chi$ and before relaxation, we can trace out the state of the electron in the donor ground orbital and restrict the time evolution to the moving electron, as demonstrated in Fig. S1. Hence, we consider {\it many-body} eigenstates $\ket{\psi_\mu}$ of the full supra-molecular Hamiltonian $H=H_\chi+H_D+H_A$, consisting of a fixed number $N+1=5$ of electrons. These can be gathered into factorized $\ket{\psi_k^N \sigma_{D/A}}$ states with $N$ electrons on the bridge and one either on the excited D orbital or on A, and $\ket{\psi_j^{N+1}}$ states with $N+1$ electrons delocalized on the bridge [Fig. \ref{fig1}-(b), bottom]. The energy-level structure of these states is sketched in Fig. \ref{fig1}-(b), with some of the $\ket{\psi_j^{N+1}}$ lying in the energy window between the 
the lowest-energy $\ket{\psi_k^N \sigma_D}$ and all the $\ket{\psi_k^N \sigma_A}$ states and hence participating in the ET. \\
To drive ET, we consider jump operators $X_D = \sum_\sigma c^\dagger_{1\sigma} c_{D \sigma} + {\rm h.c.}$ and $X_A = \sum_\sigma c^\dagger_{A\sigma} c_{4, \sigma} + {\rm h.c.}$, 
inducing spin-independent 
hopping from the donor excited orbital onto the bridge or from the bridge to the acceptor and we derive (see SI) the Redfield equation \cite{Tupkary2022} for the system density matrix $\rho$:  
\begin{equation} 
\label{eq:Redfield}
 \hbar \frac{d \rho}{d \tau} = -i [H,\rho] 
+ \Gamma \sum_{\xi=D,A} \left( Y_\xi \rho X_\xi^\dagger  - X_\xi^\dagger Y_\xi \rho  + {\rm h.c.} \right).
\end{equation}
The first term of Eq.~\eqref{eq:Redfield} describes the coherent evolution induced by the Hamiltonian $H$, while $Y_\xi = \sum_{\mu,\nu} \ket{\psi_\mu} \bra{\psi_\nu} \bra{\psi_\mu} X_\xi \ket{\psi_\nu} D_{\mu,\nu}$ and $D_{\mu,\nu}$ are proportional to the bath spectral function and to the Bose-Einstein factor at the energy gap $E_\nu-E_\mu$. In the low-temperature and wide-band limits considered hereafter $D_{\mu,\nu} = \Theta(E_\nu-E_\mu)$, $\Theta$ being the Heaviside step-function. This implies that only the subset of $\ket{\psi_j^{N+1}}$ states lying in the correct energy window [see Fig. \ref{fig1}-(b)] are involved in the ET.  

\begin{figure*}[t!]
    \centering
    \includegraphics[width=0.9\textwidth]{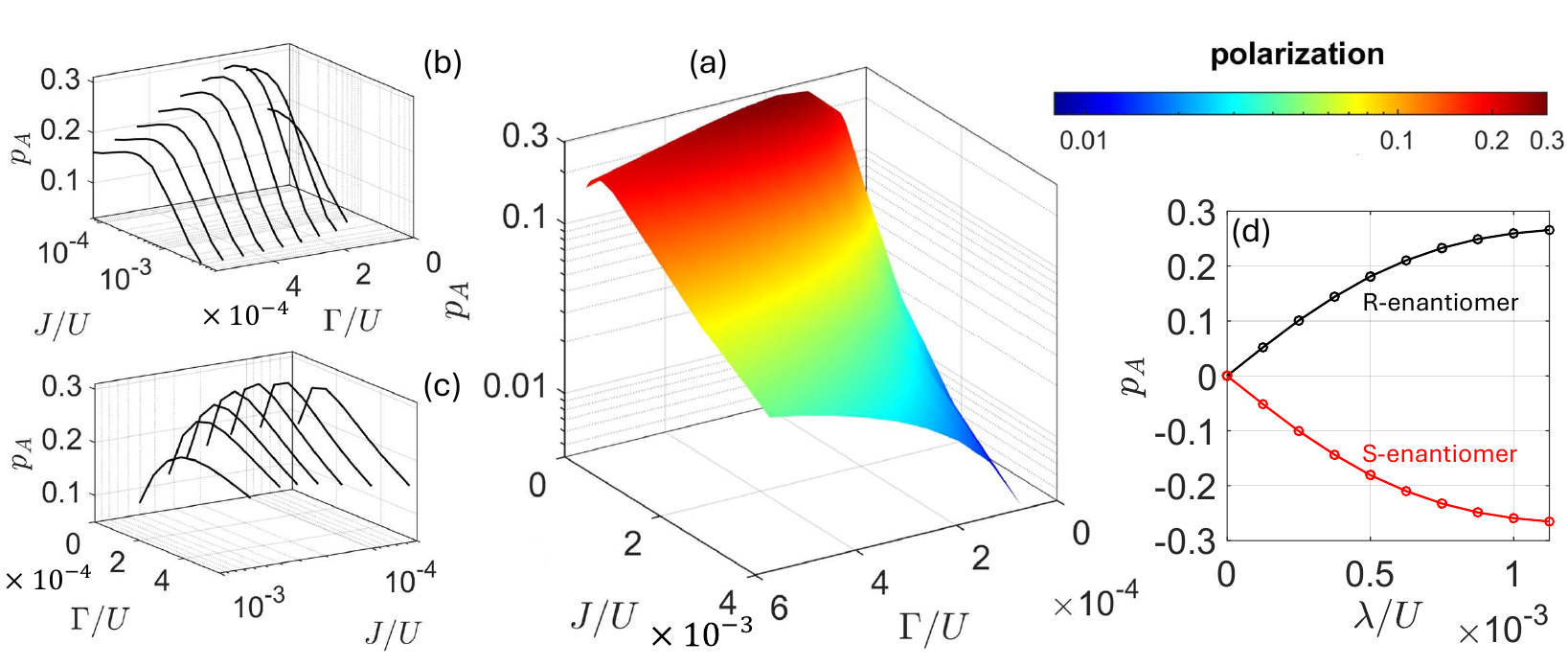}
    \caption{{\bf Dependence of spin polarization on model parameters}. (a) Acceptor polarization $p_A$ as a function of $J=4t^2/U$ 
    and of $\Gamma$ and corresponding cuts along $J$ (b) and $\Gamma$ (c) for fixed $\lambda = 6.25 \times 10^{-4} U$. (d) Dependence of $p_A$ on $\lambda$, for the two opposite enantiomers [related by the transformation $\left( \upsilon_{xi}, \upsilon_{yi}, \upsilon_{zi} \right) \rightarrow \left( -\upsilon_{xi}, \upsilon_{yi}, -\upsilon_{zi} \right)$ in Eq. \eqref{eq:Ham}], using $t/U = 0.0125$ and $\Gamma/U = 3.125 \times 10^{-4}$.  }
    \label{fig3}
\end{figure*}

An example of the simulated charge separation dynamics and of the corresponding polarization is reported in Fig. \ref{fig1}-(c,d). The system is initialized into 
$\rho(0) = \ket{\psi_0^N} \bra{\psi_0^N} \otimes (\ket{\downarrow_D} \bra{\downarrow_D} + \ket{\uparrow_D} \bra{\uparrow_D})/2$, i.e. the singlet ground state with $N$ electrons on $\chi$ and a mixture of $\ket{\uparrow}$ and $\ket{\downarrow}$ in the donor excited state, as obtained from a singlet on D, after tracing out the electron in the ground orbital.
As shown in Fig. \ref{fig1}-(c), the charge $n_i =n_{i\uparrow}+n_{i\downarrow}$ is completely transferred from D to A, after slightly populating ($N+1$)-electron states on $\chi$. 
In parallel, spin polarization $p_A = 2 S_{zA}/n_A = (n_{A \uparrow}-n_{A \downarrow})/n_A$ quickly raises on A and then stabilizes at about  20 \%  [black line in Fig. \ref{fig1}-(d)], accompanied by a negative polarization distributed on the four sites of  $\chi$ [here $S_{z,i}$ is the spin component on site $i$ along $\chi$]. 

Since we aim to focus on the effect of correlations, we report the parameters in units of the Coulomb interaction. In Fig. \ref{fig1}-(c,d) we used $t = 0.0125 \, U$, $\lambda = 6.25 \times 10^{-4} \, U$ and $\Gamma = 2.4 \times 10^{-4} \, U$. By setting $U=4$ eV, these numbers correspond to $t = 0.05$ eV, $\lambda = 2.5$ meV and $\Gamma = 1$ meV, yielding similar energy scales for $J, t, \lambda$, as discussed below.
Note that these numbers for $t$ \cite{Gagliardi2015,Fransson2019,Fransson2021,Gutierrez2013} and $\lambda$ \cite{Mujica2018,Michaeli2019,Cuniberti2020} are perfectly reasonable for many systems and in particular the small $t$ is typical of DNA \cite{Guo2012,Ratner2013,Varela2016,Cuniberti2020,Liu2021}. The resulting time-scale of ET  is in the order of 10 ps. 

{\it Insight into the spin polarization mechanism --} 
We now illustrate the mechanism 
building up a net polarization on A. To understand it, we need to focus on the many-body states $\ket{\psi_j^{N+1}}$ reported in Fig. \ref{fig2}-(a,b). The lowest energy $\ket{\psi_j^{N+1}}$ ($j=1,\ldots,32$) with a single double occupation of the bridge orbitals are organized into four blocks with different charge distribution (sketched on the left), split by  $\sim t$ [Fig. \ref{fig2}-(a)]. Each block displays a similar spin structure, consisting 
of two total spin doublets $S=1/2$ and a higher total spin $S=3/2$ quartet, split by the exchange $J = 4 t^2/U$ [Fig.~\ref{fig2}-(b)]. SOC further splits the quartet into two Kramers doublets (crosses vs. circles).  
Note that due to the energy dependent term $D_{\mu,\nu}$ in Eq. \eqref{eq:Redfield}, only some $\ket{\psi_j^{N+1}}$ states participate in the ET [we consider the lowest energy block of Fig. \ref{fig2}-(a)]. 

The ingredients contributing to the observed rise of $p_A$ are depicted in Fig. \ref{fig2}-(c-e). First {\bf (1)}, the incoherent jump operator $X_D$ generates both populations and {\it coherences} \footnote{Off-diagonal elements of $\rho$. In particular, {\it real} inter-multiplet coherences give rise to the oscillations in Fig. \ref{fig2}-(d). For illustrative purposes we consider in this section only $\lambda_z = \lambda \upsilon_{iz}$, but all the other simulations are performed with the hole $\boldsymbol{\upsilon}_i$.} [colors in panel (c)] between the different multiplets of Fig. \ref{fig2}-(b). \\ 
Then, the combined effect of $J$ (which splits different multiplets) and $\lambda$ (which mixes them) 
as sketched in Fig. \ref{fig2}-(c), yields coherent oscillations {\bf (2)} of the local polarization on different sites, and in particular of $S_{z,i=4}$ [Fig. \ref{fig2}-(d)]. Different enantiomers  yield opposite $S_{zi}$ (red/blue curves). 
Since the rise of spin polarization is triggered by the small SOC typical of organic molecules, its effect on the coherent evolution {\it can only be enhanced by the presence of a similar energy scale in the spectrum}, provided by $J$. 
The amplitude of these oscillations is proportional to the real part of the inter-multiplet coherence and increases for small $J \propto t^2/U$, i.e. strong correlations. \\
The coherent oscillations of $\langle S_{zi} \rangle$ are necessary but not sufficient to explain the polarization \emph {accumulated} on A. The last ingredient {\bf (3)} is represented 
by the incoherent terms in the Redfield Equation \eqref{eq:Redfield}. 
To shed light on this, we derive (see SI) the variation of charge on A into an elementary time step according to  Eq. \eqref{eq:Redfield}. We find   
\begin{eqnarray}
    && d  \langle n_{A\sigma}\rangle = \Tr [d \rho \; n_{A\sigma}] \\ \nonumber 
    &=& \Gamma d \tau  \Tr \left[ Y_A\rho X_A^\dagger + X_A\rho Y_A^\dagger \right]  \propto   \Tr \left[ n_{i=4,\sigma} \rho^{(N+1)}  \right], 
\label{eq:dnA}
\end{eqnarray}
i.e. the variation of $\langle n_{A\sigma}\rangle$ is proportional to the expectation value of the $n_\sigma$ on the 4th site, evaluated within the $(N+1)-$electrons subspace. Hence, $\langle S_{zA} \rangle$ is proportional to the time integral of $\langle S_{z,i=4} \rangle$ onto the $(N+1)-$electrons subspace. 
Note that $\langle S_{z,i=4} \rangle$ displays  an oscillating behavior, but it is {\it always positive} within the $(N+1)-$electrons subspace [Fig. \ref{fig2}-(e)]. Hence, its integral ($\propto \langle S_{zA} \rangle$) is a monotonic increasing function [Fig. \ref{fig2}-(f)], resulting from the interplay between coherent and incoherent dynamics  on similar time-scales \footnote{Note that the three steps {\bf (1-3)} are all inter-connected and occur simultaneously in the dynamics.}. \\
We stress that this effect would practically vanish if electron-electron correlations (responsible of the multiplet structure of the $\ket{\psi_j^{N+1}}$ states) are neglected. 

{\it Numerical simulations --} Having clarified the spin polarization mechanism, we  investigate the dependence of $p_A$ on the model  parameters. Results are summarized in Fig. \ref{fig3}. Panels (a-c) show the behavior of $p_A$ as a function of the incoherent transfer rate $\Gamma$ and of the exchange $J$, for fixed $\lambda = 6.25 \times 10^{-4} U$. We note a maximum, both as a function of $J=4t^2/U$ 
and of $\Gamma$, more evident in the cuts reported in panel b and c, respectively. This non-trivial dependence highlights the aforementioned complex interplay between coherent and incoherent dynamics. 
Remarkably, $p_A$ vanishes for vanishing correlations (i.e. large $t/U$) and reaches its maximum for  small $J$. In particular, the highest polarization $\sim 30 \%$ is achieved for $J/U = 1.44 \times 10^{-4}$ (corresponding to $t/U \approx 6.25 \times 10^{-3}$) and intermediate $\Gamma \approx 2 \times 10^{-4} U$. Note that $p_A$ also depends on the number of $\ket{\psi_j^{N+1}}$ states participating in the ET and it strongly decreases by including higher energy blocks in Fig. \ref{fig2}-(a). \\ 
Fig. \ref{fig3}-(d) also reports the increase of $|p_A|$ with $\lambda$, for both enantiomers. $p_A$ changes sign by changing enatiomer, while keeping the same absolute value. \\ 
\begin{figure}[t!]
    \centering
    \includegraphics[width=0.48\textwidth]{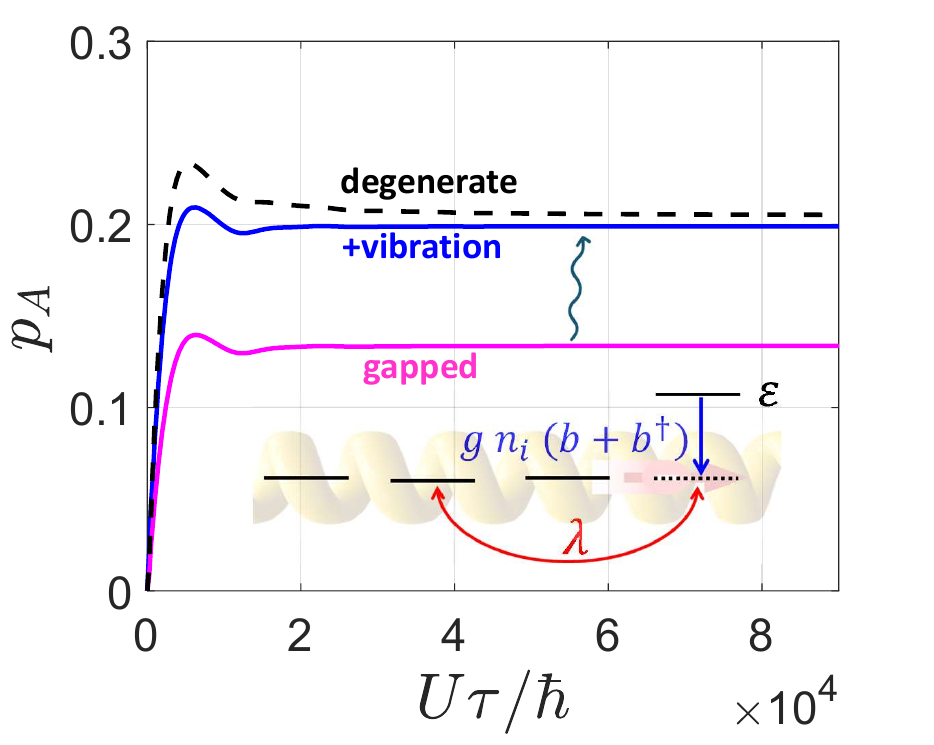}
    \caption{{\bf Non-equivalent orbitals and polarons}.     
    Effect on the acceptor polarization $p_A$ of introducing $\varepsilon$ in $H_\chi$ (magenta line) and of adding a local vibration (blue), thus approaching the degenerate situation (black dashed curve). Inset: sketch of the minimal model we consider (with the fermion-boson coupling in blue and details in the SI), with $t/U = 0.0125$, $\lambda/U = 6.25 \times 10^{-4}$, $\Gamma/U= 2.5 \times 10^{-4}$, $\varepsilon/U = 0.0375$, $\hbar \omega_0/U = 0.0125$, $g/U = 0.15$, up to 9 bosons included. }
    \label{fig4}
\end{figure}
Possible energy gaps between orbitals \cite{Ratner2013} would lead to more localized eigenstates and affect ET. We have investigated this effect and found a significant polarization retained even in presence of a sizable gap $\varepsilon \gg \lambda$ between different orbitals (see Fig. \ref{fig4}, magenta line).
Moreover, strong electron coupling with local vibrations which can occur in these systems \cite{Zeidan2008} leads to a re-normalization of the orbital energies which can effectively reduce $\varepsilon$, and hence result in polarization dynamics very similar to Fig. \ref{fig1}. 
Although an extensive study on the role of polarons is beyond the scope of this work, we show this by the simulations reported in Fig. \ref{fig4}, where an energy gap $\varepsilon$ on one site and the strong coupling with a local vibration are explicitly included (see SI).

{\it Discussion --} 
The final state obtained in the presented simulations is a superposition/mixture of various many-body eigenstates $\ket{\psi_k^N \sigma_A}$ 
and hence it is expected to undergo thermal relaxation. This effect will take place on a timescale longer than ET, but typically slower than electron paramagnetic resonance experiments \cite{Eckvahl2023}, $> 10-50$ ns. Therefore, these measurements typically probe the state of the D$-\chi-$A system with two unpaired electrons (on D and A) {\it after relaxation} of the bridge into its singlet ground state. 
Hence, we have performed simulations  starting with a singlet electron pair on D and including thermal relaxation, modeled via a rate-equation in the system eigenbasis (see SI) \footnote{For simplicity, we have only included the $\upsilon_z$ contribution to the SOC.}. \\
This includes modulation by the bath of leading one-body terms in the Hamiltonian. Moreover, to allow for 
relaxation from states with $\sum_{i \in \chi} S_{z,i} \neq 0$
onto the singlet ground state of $\chi$ \footnote{Indeed, modulation of nearest-neighbors hopping and orbital energies involve rank-0 operators on the bridge, which preserve both its total spin and its $z$ component}, we include a modulation of the coupling of the bridge end 
spins with the two spins on D and A. We thus obtain complete  relaxation of the bridge onto its singlet ground state (see Fig. S3), with its original 
polarization redistributed among D and A. The resulting spin state of the DA radical pair is still spin polarized with opposite polarization on D and A, as observed experimentally. \\ 
Another point concerns the dependence of our results on the length of the chiral bridge. Two different situations should be distinguished. In the first case, ET could take place in only two incoherent steps (as considered above) from D to $\chi$ and from $\chi$ to A, interleaved by the coherent evolution of the bridge many-body states. In this case a longer bridge with similar parameters $J \sim \lambda$ 
will exhibit a qualitatively similar coherent evolution.
We have checked this by numerically computing the dynamics for $N=6$ and found a spin polarization comparable to that obtained with $N=4$ using the same parameters (see  Fig. S4). \\
A different possibility is that ET occurs as a sequential multi-step incoherent process through different parts of a longer bridge. In that case, the process can be approximately described as a concatenation of steps 
in which the final spin state 
becomes the initial state of the next one. As a result, spin polarization can accumulate in each step. For instance, re-starting from $p_A=0.20$ we get $0.27$ and $0.29$ in the following two steps as reported in Fig. S4. These numbers depend on the choice of parameters, but indicate that in a sequential multi-step ET the spin polarization can increase with the length of the bridge, as observed in conductance measurement on thick films of DNA and oligopeptides \cite{Bloom2024}. 

To conclude, we have reported the first microscopic model for CISS in electron transfer through chiral molecules, with the explicit inclusion of the bridge degrees of freedom that play an active role in the electron spin polarization. 
Based on experimental evidence of spin polarization even on short chiral chains, we have built a minimal model system on a limited number of sites and we have pinpointed the essential ingredients for achieving a sizable spin polarization even in presence of a small spin-orbit coupling. These are: (i) strong electron-electron correlations giving rise to many-body states split by the exchange interaction; (ii) an interplay of coherent (exchange + spin-orbit) dynamics on the chiral bridge and incoherent hopping from the donor or onto the acceptor; (iii) relaxation of the chiral bridge to establish the long term polarization observed experimentally.\\ 



{\bf Acknowledgments} \\
We warmly thank A. Painelli and A. Phan Huu for very fruitful and stimulating discussions. 
The work was funded by the Horizon Europe Programme within the ERC-Synergy project CASTLE (proj. n. 101071533). 
Views and opinions expressed are however those of the author(s) only and do not necessarily reflect those of the European Union or the European Commission. Neither the European Union nor the granting authority can be held responsible for them. \\ 
M.M. acknowledges funding from the European Union – NextGenerationEU under the National Recovery and Resilience Plan (NRRP), Mission 4 Component 1 Investment 3.4 and 4.1. Decree by the Italian Ministry n. 351/2022 CUP D92B22000530005.

\end{document}